\documentclass[conference,a4paper]{IEEEtran}
\usepackage{amsmath}
\usepackage{amsbsy}
\usepackage{amsfonts}
\usepackage{amssymb,hhline}
\usepackage{epsfig}
\usepackage{graphicx}

\usepackage{url}
\usepackage{amsthm}
\usepackage[ruled,boxed,boxruled]{algorithm2e}

\ifCLASSINFOpdf
\else
\fi
\DeclareRobustCommand*{\IEEEauthorrefmark}[1]{\raisebox{0pt}[0pt][0pt]{\textsuperscript{\footnotesize #1}}}

\newcommand{\paino}{\sl}

\newcommand{\bom}[1]{\boldsymbol{#1}}
\newcommand{\bo}[1]{\mathbf{#1}}

\newcommand{\s}{\bom \beta}  

\renewcommand{\a}{\bo a}  
\newcommand{\es}{\beta}      
\newcommand{\y}{\bo y} 
 
\newcommand{\e}{\bo r}  
\renewcommand{\b}{\bo b} 
\renewcommand{\r}{\bo r} 

\newcommand{\mm}{\bom \Phi}   

\newcommand{\A}{\bo A}

\newcommand{\iidsim}{\overset{iid}{\sim}}

\newcommand{\sig}{\sigma}

\newcommand{\eps}{\bo e} 
\newcommand{\eeps}{\varepsilon} 
\newcommand{\al}{\alpha} 
 
\newcommand{\bth}{\bom \theta} 

\newcommand{\be}{\alpha} 
\newcommand{\im}{\jmath} 
\newcommand{\ssgn}{\mathrm{sign}} 

\newcommand{\hop}{\mathsf{H}}

\newcommand{\R}{\mathbb{R}}    
\newcommand{\C}{\mathbb{C}}    

\newcommand{\expec}{\mathbb{E}}    

\newcommand{\pr}{\partial}
\newcommand{\lam}{\lambda}

\newcommand{\beq}{\begin{equation}}
\newcommand{\eeq}{\end{equation}}
\newcommand{\bmat}{\begin{pmatrix}}
\newcommand{\emat}{\end{pmatrix}}
\newcommand{\beqa}{\begin{eqnarray}}
\newcommand{\eeqa}{\end{eqnarray}}

\newcommand{\mb}{\bom \phi}    
\newcommand{\ndim}{n}             
\newcommand{\pdim}{p}             
\newcommand{\kdim}{k}             

\newcommand{\veps}{\bom \varepsilon}

\begin{document}
%
\title{Direction of arrival estimation using robust complex Lasso}

\author{\IEEEauthorblockN{
Esa Ollila
}                                     
\IEEEauthorblockA{Department of Signal Processing and Acoustics, 
Aalto University \\ P.O.Box 13000,  FI-00076 Aalto, Finland} 
}



\maketitle

\begin{abstract}
The Lasso (Least Absolute Shrinkage and Selection Operator) has been a popular technique for simultaneous linear regression estimation and variable selection. 
  In this paper, we propose a new novel approach for robust Lasso that follows the spirit of $M$-estimation. 
We define $M$-Lasso estimates of regression and scale as solutions to generalized zero subgradient equations. 
 Another unique feature of this paper is that we consider  complex-valued measurements and regression parameters, 
which requires careful mathematical characterization of the problem. 
An explicit and  efficient algorithm for computing the $M$-Lasso solution is proposed that 
has  comparable computational complexity as state-of-the-art algorithm for  computing the Lasso solution. Usefulness of the $M$-Lasso method is illustrated for  direction-of-arrival (DoA) estimation with sensor arrays in a single snapshot case.  

\end{abstract}

{\smallskip \keywords Compressive sensing, beamforming,  DoA estimation, Lasso, sparsity}

%
\IEEEpeerreviewmaketitle

\vspace{7pt}
\section{Introduction}

We consider  the complex-valued linear model 
$\y = \mm \s + \veps$,    
where   
 $\mm$ is a known $\ndim \times \pdim$ complex-valued  measurement matrix (or matrix of predictors), $\s=(\es_1,\ldots,\es_\pdim)^\top$
is the unknown vector of complex-valued regression coefficients (or system parameters) 
and $\veps \in \C^{\ndim}$ denotes the additive noise.  For ease of exposition, we consider the centered linear model (i.e., we assume that the intercept is equal to zero). 
The primary interest is to estimate the unknown parameter $\s$ 
given $\y\in \C^{\ndim}$  and $\mm \in \C^{\ndim \times \pdim}$. 
 However, in many practical applications,  the linear system is {\paino underdetermined}  ($\pdim > \ndim$) or $p \approx n$ and the least squares estimate (LSE)  does not have a unique solution 
 or is subject to a very high variance. Furthermore, for large number of predictors, we would like to identify the ones that exhibit the strongest effects. Hence we wish to find a {\paino sparse solution} $\hat \s$,  which sets weights for irrelevant predictors  equal to $0$. 
 In these cases one needs to regularize the regression                           
coefficients (i.e., to control how large they can grow). 
 Another problem with the LSE arises when there are outliers or the noise follows a heavy-tailed non-Gaussian distribution. Then robust estimation \cite{huber:1981} 
 is upmost importance for reliable estimation of the unknown parameters.

The  complex version of the popular Lasso  \cite{LASSO:1996} solves
an $\ell_1$-penalized LS regression problem,  
\beq \label{eq:penfunc} 
\min_{\s} \| \y - \mm \s \|^2_2  +  2 \lam  \| \s\|_1 
\eeq 
where $\lambda >0 $ is the  shrinkage (penalty) parameter. 
As $\lambda \in (0, \infty)$  varies,  the solution $\hat \s_\lambda$ 
traces out a path in $\C^\pdim$, with $\hat \s_{\lambda\to0}$ then corresponding to the conventional LSE.   We refer the reader to  \cite{hastie2015statistical} for a comprehensive account on Lasso.  The larger the value of $\lambda$  the greater is the amount of shrinkage for the coefficients (some of which can be shrunk all the way to zero).

Robust Lasso is needed in case of heavy-tailed errors or severe outliers. 
A popular focus  in the literature for obtaining robust Lasso estimates is to use a robust criterion 
in place of the  least squares (LS) criterion.  Most robust loss functions  require a preliminary estimate of the scale of the error terms.  
An accurate estimate of scale is difficult to obtain since  significant predictors are unknown ($\s$ is sparse and possibly $\ndim < \pdim$). Therefore  a joint estimation of regression and scale becomes a necessity.  In this paper, we propose a new  approach for robust Lasso that follows the spirit of $M$-estimation. 
We define $M$-Lasso estimates of regression and scale as solutions to generalized zero subgradient equations which  are based on general loss function.  
These equations are a sufficient and necessary condition of a solution  to the Lasso problem \eqref{eq:penfunc} given that the loss function is the LS-loss.   A unique feature of this paper is that we consider  complex-valued measurements and regression parameters.  This requires careful mathematical characterization of the problem and proper tools from complex function theory.   A simple 
and efficient algorithm for computing the $M$-Lasso solution is also developed. 

We illustrate how the proposed $M$-Lasso can be used for DoA estimation of source signals  using sensor arrays when only a  single snapshot  is available.  
Indeed sparse regression approaches for DoA estimation has been an active research field; see 
\cite{malioutov_etal:2005,fortunati_etal:2014,xenaki2014compressive,gerstoft_etal:2015} and references therein.  
Our examples illustrate that $M$-Lasso based on Huber loss function has similar performance in DoA finding as  Lasso  \eqref{eq:penfunc} in  complex Gaussian noise, but superior performance in heavy-tailed non-Gaussian noise or in face of outliers. 

Let us offer a brief outline of the paper.  Robust loss functions and their properties in complex-valued case are outlined in Section~\ref{sec:loss}.  Also the notion of pseudo-residual vector 
is introduced which will be elemental in our developments.  
In Section~\ref{sec:Mlasso}, we recall the zero  subgradient estimating equations for Lasso solution and then define $M$-Lasso estimates of regression and scale as solutions to generalized subgradient equations.  A highly efficient algorithm for computing the $M$-Lasso estimates is also proposed. Finally, we consider the direction finding application with sensor arrays in Section~\ref{sec:doa}. Section~\ref{sec:concl} concludes.  


{\it Notations}. The vector space  $\C^{\ndim}$ is
equipped with the usual Hermitian inner product, $\langle \a, \b \rangle = \a^\hop \b$,  where $(\cdot)^\hop=[(\cdot)^*]^\top$ denotes the Hermitian  (complex conjugate) transpose.   
This induces the conventional (Hermitian) $\ell_2$-norm $\| \a \|_2 = \sqrt{\a^\hop \a}$.   The $\ell_1$-norm is the defined 
as $\| \a\|_1 = \sum_{i=1}^\ndim |a_i |$, where $| a | = \sqrt{a^* a} = \sqrt{a_R^2 + a_I^2} $ denotes the modulus of a complex number 
$a=a_R + \im a_I$.  For a matrix $\bo A \in \C^{\ndim \times p} $, we denote by $\bo a_i \in \C^\ndim$ its $i^{th}$ column vector 
and $\a_{(i)} \in \C^{\pdim}$ denotes the Hermitian transpose of its $i^{th}$ row vector. Hence, this means that  we may write  the measurement matrix $\mm \in \C^{\ndim \times \pdim}$ as $\mm= \bmat \mb_{1} & \cdots & \mb_{\pdim}\emat = \bmat \mb_{(1)} & \cdots & \mb_{(\ndim)}\emat^\hop$.


\vspace{7pt}
\section{Robust loss functions and pseudo-residuals} \label{sec:loss}

Suppose that the error terms $\eeps_{i}$ are i.i.d. continuous random variables from a circular distribution \cite{ollila_etal:2011} with p.d.f. $f(e)=(1/\sigma)f_0(e/\sig)$, where 
$f_0(e)$ denotes the standard form of the density and $\sigma>0$ is the scale parameter.   Robust loss functions commonly require knowledge of $\sig$ in order to properly 
downweight outlying observations.  Hence the unknown scale $\sig$ needs to be estimated jointly with the regression coefficient as a preliminary robust scale estimate is generally not available.  

We adopt the definition of {\paino loss function} to complex-valued case from \cite{ollila:2015b}. Namely, we call $\rho: \C \to \R^+_0$
a  loss function if it is circularly symmetric,  $\R$-differentiable convex function 
which satisfies  $\rho(0)=0$. %
 Due to circularity assumption (implying that $\rho(e^{\im \theta} x)=\rho(x) \forall \, \theta \in \R$) it follows that 
$\rho (x)=\rho_0(|x|)$  
for some $\rho_0 : \R^+_0 \to  \R^+_0$.  
This illustrates  that $\rho$ is not $\C$-differentiable (i.e., holomorphic function) since 
only functions that are {\it both} holomorphic {\it and} real-valued are constants. 
The complex derivative \cite{eriksson_etal:2010}  of $\rho$ w.r.t. $x^*=(x_R + \im x_I)^*$ is  
 \begin{align*} 
 \psi(x) &= \frac{\partial}{\partial x^*}\rho(x) = \frac 12 \left(\frac{\pr \rho}{\pr x_R} + \im  \frac{\pr \rho}{\pr x_I} \right) = \frac 1 2 \rho_0'( |x|) \ssgn(x) ,
\end{align*} 
where 
\[
\ssgn(e)= \begin{cases}   e/| e |, &\mbox{for  $e \neq  0$} \\     0 ,&\mbox{for $e = 0$} \end{cases} 
\]
is the  complex  signum function  and $\rho_0'$ denotes the real derivative of the real-valued function $\rho_0$.  
Function $\psi: \C \to \C$ will be referred in the sequel   as {\paino score function}.

For obtaining robust estimates, the utilized loss function $\rho(e)$ should assign smaller weights for large errors $e$ than the {\paino LS (or $\ell_2$-)loss} $\rho(e)=|e|^2$. 
One most commonly used robust loss function is due to Huber   \cite{huber1964robust}. In the complex-valued case, {\paino Huber's loss function}  
can be defined as  follows \cite{ollila:2015b}: 
\beq \label{eq:huber} 
\rho_{H,c}(e) =  
\begin{cases}  |e|^2, &\mbox{for  $|e| \leq c$} \\   2c |e| - c^2, &\mbox{for  $|e| > c$}, \end{cases}
\eeq 
where  $c$ is a  user-defined {\paino threshold} that influences the degree of robustness 
and efficiency of the method.  Hence similar to the real-valued case, Huber's loss function 
is a hybrid of $\ell_2$ and $\ell_1$ loss functions $\rho(e)=|e|^2$ and $\rho(e)=|e|$, respectively, using $\ell_2$-loss for relatively small errors and $\ell_1$-loss for relatively large errors.  Moreover, it is convex and  verifies the conditions imposed on the loss function ($\R$-differentiability and circular symmetry). 
Huber's score function becomes
\[
\psi_{H,c}(e) = \begin{cases} e, &\mbox{for  $|e| \leq c$} \\  c \,  \ssgn(e), &\mbox{for $|e|>c$}\end{cases}.
\]
Note that Huber's $\psi$ is a winsorizing (clipping) function: the smaller the $c$, the more clipping is actioned on the residuals. 

Now recall that in robust regression, the loss function acts on standardized residual vector $\e/\sig$, where  $\e\equiv \e(\s)=\y-\mm \s$ denotes the residual vector for  some candidate  $\s \in \C^\pdim$ of regression coefficient vector and $\sig$ is the scale. The  loss function then defines a  {\paino pseudo-residual}, defined as    
\beq  \label{eq:pseudo_residual} 
\e_\psi \equiv \e_\psi(\s,\sig)= \psi\! \left( \frac{\y-\mm  \s}{\sigma}\right)  \sig
\eeq 
where   $\psi$-function acts coordinate-wise to vector $\e/\sig$, so $[\psi(\e/\sig)]_i= \psi(r_i/\sig)$.  
Some remarks of definition \eqref{eq:pseudo_residual} are in order. First, note that if $\rho(\cdot)$ is  the conventional LS-loss, $\rho(e)=|e|^2$,  then $\psi(e)=e$, and $\e_\psi$ is equal with  the conventional residual vector,  so $\e_\psi = \y-\mm \s=\e$. Second, for Huber's loss function, pseudo-residual vector $\e_\psi$  has $i^{th}$ element equal to  
$r_i$ if $| r_i| < c \sig$
and  equal to $(c \sig) \ssgn(r_i)$ otherwise. In other words, residuals that are farther  apart from zero than $c$ times the scale $\sigma$ are trimmed (downweighted). This is the underlying principle for robustness of Huber's loss function. Third, note that the multiplier $\sig$ in  \eqref{eq:pseudo_residual}   is essential in bringing the residuals back to
 the original scale of the data. 
 
 Since $\sig$ is unknown in practise, robust loss function  $\rho(e)$, such as Huber's loss above,  require a preliminary robust scale estimate  $\hat \sig$ in order to determine  if $e$ should be downweighted or not. In sparse regression problems, obtaining such an estimate is more difficult task than in the conventional regression problem since now the significant predictors (columns of $\mm$) are not known ($\s$ is sparse and possibly $\ndim<\pdim$). Suppose that a preliminary robust regression estimate $\hat \s_{init}$ can be computed ($\ndim > \pdim$ case) and then used to compute  a robust scale statistic $\hat \sig$ (e.g., median absolute deviation) based on  $\hat \r=\y - \mm \hat \s_{init}$.  Such a scale estimate is biased and can significantly underestimate the true $\sig$ due to overfitting when the number of predictors $\pdim$ is large.   This then implies too severe downweighting.    
So robust loss function and  underestimated $\hat \sig$ results in  pseudo-residuals which can severely downweight 'good' residuals (not just outliers). Similarly, all residuals  
can be left intact if $\hat \sig$ is an overestimate (too large).

\section{Robust complex $M$-Lasso} \label{sec:Mlasso}

The  earlier approaches for robust Lasso are based on an idea of adding 
an $\ell_1$-penalty to a robust criterion function since the LS criterion function, $
J_{\ell_2}(\s)=\| \y - \mm \s \|^2_2$, is sensitive to outliers.  For example,   \cite{wang_etal:2007} utilize 
the least absolute deviation (LAD) criterion, $J_{\ell_1}(\s)=\| \y - \mm \s \|_1$, LTS-Lasso  of \cite{alfons_etal:2013} 
is based on the  least trimmed squares (LTS) criterion, whereas \cite{owen2007robust} utilized Huber's criterion function $Q(\s,\sig)$ in  \eqref{eq:Q}. 

Our approach is different from these earlier approaches.   Namely, we define Lasso estimator as a solution to generalized 
zero subgradient equations that is based on general loss function $\rho(e)$. 
Our approach follows the spirit of $M$-estimation \cite{huber1964robust,huber:1981}  where the principal idea is to define an estimator as a solution to 
generalized maximum likelihood (ML-)estimating equations.    

We start by recalling the zero 
subgradient equation for complex-valued Lasso  problem \eqref{eq:penfunc}. 
Note that the utilized  LS criterion function $J_{\ell_2}(\s)$ in Lasso problem \eqref{eq:penfunc} 
is convex (in fact strictly convex if $\ndim > \pdim$) and 
$\R$-differentiable but the $\ell_1$-penalty  function $\| \s\|_1$ is not $\R$-differentiable at a point where at least one coordinate $\es_j$ is zero.  
However, we can resort to generalization of notion of gradient applicable for convex functions, called the {\paino subdifferential} \cite{boyd2004convex}. 
For a complex function $f:\C^\pdim \to \R$ we can define subdifferential at a point $\s$ as   
\begin{align*} 
\pr f(\s)= \{ \bo z \in \C^\pdim : f(\s') \geq  f(\s) + 2 &\mathrm{Re}( \langle \bo z, \s' - \s \rangle) \\ 
&\quad \mbox{for all } \s' \in \C^\pdim  \} .
\end{align*} 
Any element $\bo z \in \pr f(\s)$ is then called a {\paino subgradient} of $f$ at $\s$. 
The subdifferential of  the modulus  $| \es_j |$  is 
\[
\pr | \es_j | =  \begin{cases} \frac 1 2 \ssgn(\es_j), &\mbox{for $\es_j \neq 0 $} \\   \frac 1 2s \,   &\mbox{for $\es_j = 0 $}\end{cases}
\] 
where $s$ is some complex number verifying $| s | \leq 1$. Thus subdifferential  of $|\es_j|$ is the usual complex  derivative when  $\es_j \neq 0 $, i.e., 
 $\pr | \es_j| = \frac{\pr}{\pr \es_j^*} | \es_j| $ for $\es_j \neq 0 $. 
Then a necessary and sufficient condition for a solution to the Lasso problem \eqref{eq:penfunc}  
is that 
$
\pr (J_{\ell_2}(\s)  +  2 \lam  \| \s \|_1) \in \bo  0$
which gives zero subgradient equation
\beq 
-\mb^\hop_{i} \big(  \y-\mm \hat \s \big) + \lam \hat s_j  = 0 \quad \mbox{for } j = 1,\ldots,\pdim \label{eq:LSestim1}
\eeq 
where $\hat s_j$ is 2 times  an element of the subdifferential  of $|\es_j|$ evaluated at $\hat \es_j$, i.e.,   
equal to $\ssgn(\hat \es_j)$ if $\hat \es_j \neq 0$ and some complex number lying inside the unit complex circle otherwise.
Given the Lasso solution $\hat \s$, the natural scale estimate $\hat \sig^2$ is then 
\beq \label{eq:scale} 
\hat \sigma^2 = \frac{1}{n} \sum_{i=1}^\ndim \big| y_i - \mb_{(i)}^\hop \hat \s \big|^2  = \frac{1}{n} \| \hat \e \|^2_2  
\eeq
where $\hat \e=\y - \mm \hat \s$ denote the residual vector at the solution. 


Given the considerations in Section~\ref{sec:loss}, it appears wise to estimate the unknown parameters $\s \in \C^\pdim$ and $\sig>0$ jointly.    
Thus we seek for a pair $(\hat \s,\hat \sig)$ which verify the generalized  (zero subgradient) estimating equations,  which we refer to as {\paino Lasso $M$-estimating equations},   
of the form 
\begin{align} 
-\mb^\hop_{i}  \e_\psi(\hat \s,\hat \sig)   + \lam \hat s_j  &= 0 \quad \mbox{for } j = 1,\ldots,\pdim  \label{eq:estim1} \\ 
  \al \ndim -  \sum_{i=1}^\ndim \chi\! \left( \frac{|y_i- \mb_{(i)}^\hop \hat \s|}{\hat \sig} \right) &= 0   \label{eq:estim2} 
\end{align}
where   $\al>0$ is a fixed scaling factor (described later) and function $\chi: \R_0^+ \to  \R_0^+$ is defined as
\beq \label{eq:chi}
\chi(t)=\rho_0'(t) t - \rho_0(t) .
\eeq 
Recall that $\rho(x)=\rho_0(|x|)$. To simplify notation we write  the pseudo-residual vector $\e_\psi(\hat \s,\hat \sig)$ in \eqref{eq:estim1} as $\hat \e_\psi$. 

Some remarks of  this definition are in order. First, consider the  conventional choice, i.e., the LS-loss $\rho(e)=|e|^2$. 
 In this case, $\hat \e_\psi= \hat \e$,  so  \eqref{eq:estim1} reduces to \eqref{eq:LSestim1}. Furthermore,   
since $\rho_0(t)=t^2$ and $\rho_0'(t)=2t$, the $\chi$-function in \eqref{eq:chi} is $\chi(t)=t^2$, and \eqref{eq:estim2} reduces 
 to \eqref{eq:scale}. In other words, for LS-loss, the $M$-Lasso solution $(\hat \s,\hat \sig)$  to \eqref{eq:estim1}-\eqref{eq:estim2} 
is the conventional Lasso estimate (so $\hat \s$ is a solution to \eqref{eq:penfunc}) and $\hat \sig$ equals the standard scale statistic in \eqref{eq:scale}.  
Second, if $\lam=0$ (so no penalization and $\ndim > \pdim$), then the solution to  \eqref{eq:estim1}  and \eqref{eq:estim2} is the unique solution to the 
convex optimization problem
\begin{align} \label{eq:Q}
\arg \min_{\s, \sig} \bigg \{ Q(\s,\sig)  =   \al \ndim \sig  +  \sum_{i=1}^\ndim \rho \bigg(  \frac{y_{i} - \mb_{(i)}^\hop \s}{\sig} \bigg) \sig  \bigg\} . 
\end{align} 
Important feature of the objective function $Q(\s,\sig)$ above  is that it  is  jointly convex in $(\s,\sig)$ given that $\rho$ is convex.  
In other words, for $\lam=0$ (and $\ndim > \pdim$),    equations \eqref{eq:estim1}  and \eqref{eq:estim2} are  necessary and sufficient condition for a solution to problem \eqref{eq:Q}. This objective function 
was originally studied by Huber \cite{huber:1981} in the real-valued case. 
Lasso penalized Huber's criterion was considered by Owen \cite{owen2007robust} and $\ell_0$-penalization   in real-valued and complex-valued case in \cite{ollila_etal:2014,ollila:2015b}, respectively.

Next we note that  \eqref{eq:estim1}  can be written after recalling the definition  \eqref{eq:pseudo_residual}  more compactly as 
$
\langle \mb_j , \hat \r_\psi \rangle =  \lam  \hat s$ 
for $j=1,\ldots,\pdim.
$
This mean that (after taking modulus of both sides) the following holds 
\begin{align} 
| \langle \mb_j , \hat \r_\psi \rangle | &= \lam, \quad \mbox{ if }  \hat \es_j  \neq 0  \label{eq:lam1}  \\
 | \langle \mb_j , \hat \r_\psi \rangle | &\leq\lam, \quad \mbox{ if }  \hat \es_j  = 0  \label{eq:lam2} 
\end{align} 
i.e., whenever a component, say $\hat \es_j$,  of $\hat \s$ becomes non-zero, the
corresponding absolute correlation  between the pseudo-residual $\hat \r_\psi$ and column $\mb_j$ of $\mm$, $|\langle \mb_j , \hat \r_\psi \rangle|$,   
meets the boundary $\lam$ in magnitude,  where $\lam>0$ is the penalty parameter. 
This is well-known property of Lasso; see e.g., \cite{hastie2015statistical} or \cite{gerstoft_etal:2015} in the complex-valued case. This property is then fulfilled by 
$M$-Lasso estimates by definition.  In the real-valued case,   \cite{owen2007robust} considered minimization of penalized Huber's criterion $ Q_\lam(\s,\sig)=  Q(\s,\sig)+ \lam \| \s\|_1$. The solution of $\min_{\s,\sig} Q_\lam(\s,\sig)$, however, is  different  
from solutions to \eqref{eq:estim1}-\eqref{eq:estim2}. This can be verified by noting that the zero subgradient equation 
$\pr_{\s} Q_\lam(\s,\sig) = \bo 0$   is different from \eqref{eq:estim1}.  This also means that solution for penalized Huber's criterion 
based on  LS-loss function $\rho(e)= |e|^2$ is not the Lasso solution \eqref{eq:penfunc}. This is somewhat counterintuitive. This equivalence with Lasso and $M$-Lasso for LS-loss, however,  holds.  

The scaling factor $\al$  in  \eqref{eq:estim2}  is chosen so that the obtained scale estimate  $\hat \sig$ 
is Fisher-consistent for the unknown scale $\sigma$ when  $\{\eeps_{i}\}_{i=1}^\ndim \iidsim \C \mathcal N(0,\sig^2)$.  Due to  \eqref{eq:estim2},  
it is chosen so that 
$\be =  \expec[\chi(e)]$,  when 
$e\sim \C \mathcal N(0,1)$,   
holds. 
For many loss functions, $\be$ can be computed in closed-form.   
For example, for Huber's function \eqref{eq:huber}  the $\chi$-function in \eqref{eq:chi} becomes  
\beq \label{eq:huber_chi} 
\chi_{H,c}(|e|)=  |\psi_{H,c}(e) |^2 
= \begin{cases}  |e|^2, &\mbox{for  $|e| \leq c$} \\    c^2, &\mbox{for  $|e| > c$} \end{cases} .
\eeq 
In this case the estimating equation   \eqref{eq:estim2} can be written as 
\[
 \sum_{i=1}    \left|\psi_{H,c}\bigg( \frac{y_i- \mb_{(i)}^\hop \hat \s}{\hat \sig} \bigg)  \hat \sig \right|^2 =  \hat \sig^2 n \alpha  \Leftrightarrow  \hat \sigma^2 =\frac{1}{\ndim \alpha}  \| \hat  \e_\psi \|^2  
\]
where $\hat \e_\psi = \e_\psi(\hat \s,\hat \sig)$. 
The consistency factor $\be=\be(c)$ can be computed in closed-form as 
\begin{align} \label{eq:be} 
\be     &= c^2(1-F_{\chi^2_2}(2c^2)) + F_{\chi^2_4}(2c^2) .
\end{align}
Note that $\al$ depends on the threshold $c$. We will choose threshold $c$ as 
$
c^2 =  (1/2)F^{-1}_{\chi^2_2}(q)
$
for $q\in (0,1)$. See \cite{ollila:2015b}.  

Next we propose an explicit and  efficient algorithm for computing the $M$-Lasso solution.  The algorithm follows the idea of state-of-the-art algorithm, the cyclic coordinate descent (CCD) \cite{friedman_etal:2007}, for computing the Lasso solution. Our algorithm is a generalization of it in two aspects. First, it adapts it to the complex-valued case 
and second, it generalizes the algorithm to the robust $M$-estimation scenario.  
First recall that CCD algorithm repeatedly cycles through the predictors updating one parameter  (coordinate) $\es_j$ at a time ($j=1,\ldots,\pdim$) while keeping  others fixed at their current iterate values.  At  $j$th step, the update for $\hat \es_j$ is obtained by soft-thresholding a conventional coordinate descent update $\hat \es_j +   \langle \mb_j, \hat \e \rangle$, where $\hat \e$ denotes the residual vector $\hat \e=\e(\hat \s)$ at current estimate $\hat \s$. For $M$-Lasso, similar updates are performed but $\hat \e$  replaced  by pseudo-residual vector $\hat \e_\psi$ and the update for scale  is obtained prior to cycling through the coefficients. 
The $M$-Lasso algorithm proceeds as follows: 
\begin{enumerate} 
\item  Update the scale  $\hat \sig^2 \leftarrow \dfrac{\hat \sig^2}{\al \ndim} {\displaystyle \sum_{i=1}^\ndim \chi\bigg( \frac{y_i- \mb_{(i)}^\hop \hat \s  }{\hat \sig}  \bigg)}$
\item  For $j=1,\ldots, \pdim$ do 
\begin{itemize} 
\item[a)] Update the pseudoresidual: $\hat \e_\psi \leftarrow \psi \bigg( \dfrac{\y- \mm \hat \s}{\hat \sig} \bigg) \hat \sig$
\item[b)] Update the coefficient: $\hat \es_j \leftarrow S_\lam \big( \hat \es_j  + \langle \mb_j, \hat \e_\psi \rangle \big)$ 
\end{itemize} 
\item Repeat Steps 1 and 2 until convergence
\end{enumerate} 
Above $S_\lam(x) = \ssgn(x) ( | x | - \lam)_+$, $x \in \C$, is the complex soft-thresholding operator and $(t)_+$ denotes the positive part of $t \in \R$:  $(t)_+=t$ if $t>0$ and $0$ otherwise. The $M$-Lasso algorithm  has  comparable computational complexity as state-of-the art algorithm for computing the 
Lasso solution \eqref{eq:penfunc}. 

\section{Single snapshot DoA estimation} \label{sec:doa}

We consider uniform linear array (ULA) consisting of $\ndim$ sensors with half a wavelength inter-element spacing 
that receives $\kdim$ narrowband incoherent farfield plane-wave sources from a point source ($\ndim>\kdim$). 
At discrete time $t$,  the array output (called {\paino snapshot})  $\y \in \C^\ndim$ is a weighted linear 
combination of the signal waveforms $\bo s = (s_1, \ldots,s_\kdim)^\top$ 
corrupted by additive noise $\eps \in \C^\ndim$,  
$\y(t) = \A(\bom \theta) \bo s + \eps$,  
where  $\A=\bo A(\bom \theta)$ is the $\ndim \times \kdim$ {\paino steering matrix} para\-met\-rized by the vector $\bom \theta=(\theta_1,\ldots,\theta_\kdim)^\top$ 
of  (distinct) unknown direction-of-arrivals (DoA's)  of the sources. We assume that only a single snapshot is available. 
Each column vector $\a(\theta_i)$, called  the {\paino steering vector}, 
represents a point in known array manifold, 
$ \a(\theta) = \frac{1}{\sqrt{\pdim}} (1,e^{-\im \pi  \sin(\theta)},\cdots,e^{-\im \pi (\ndim-1) \sin(\theta) })^\top$. 
The objective of sensor array source localization is to find the DoA's of the sources, i.e.,  to identify the steering matrix $\A(\bth)$ 
parametrized by $\bth$. 

As in \cite{malioutov_etal:2005},   we cast the source localization problem as a sparse regression problem.
We construct   an angular grid (look directions of interest) of size $ \pdim \gg \kdim$, $[\theta] = \{ \theta_{(i)} \in [-\pi/2, \pi/2) \ :  \   \theta_{(1)} >  \cdots > \theta_{(\pdim)} \}$.  If $[\theta]$ contains the true DoA's $\theta_i$, $i=1,\ldots,\kdim$, then 
the snapshot follows sparse linear regression model,  
$
\y = \mm \s + \veps$, 
where the measurement matrix $\mm \in \C^{\ndim \times \pdim}$ has as its columns the steering vectors at considered 
look directions, i.e., $\mb_i = \a(\theta_{(i)})$. Thus identifying the true DoA's is equivalent to identifying the non-zero elements of $\es_j$.  Thus ($M$-)Lasso estimation becomes  necessary since  often $\pdim > \ndim$ and the LSE does not provide sparse solutions.  Note also that even if $[\theta]$ does not contain the true DoA's but has reasonably fine grid, one can identify good estimates of true DoA's as locations in the angular grid corresponding to $\kdim$ largest  coefficients  of $M$-Lasso solution (given $\y$ and $\mm$)  $\hat \s_\lam$, where $\lam${}  is such that the solution has $\geq 3$ nonzero coefficients.  
If  the number of sources $\kdim$ is known, then 
more obvious approach is to obtain the $M$-Lasso estimate $\hat \s_\lam$  for a penalty  parameter $\lam$ that results in   $\kdim$-nonzero elements. 
Let us denote the largest $\lam$  value that produces the desired $\kdim$ non-zero coefficients by $\lam^*$. The locations of the nonzero coefficients of  
$\hat \s_{\lam^\star}$  in the angular grid  $[\theta]$ then give natural $M$-Lasso DoA estimates. 

The simulation set-up is described next. The ULA receives $\kdim=3$ sources
at DoA's  $\theta_1=-5$, $\theta_2=0$ and $\theta_3=20$ degrees and the noise $\veps$ has i.i.d. elements from $\C \mathcal N(0,\sigma^2)$ distribution.  
The   amplitudes of the sources are  $|s_1|=1$, $|s_2|=0.6$ and $|s_3|= 0.2$ and the noise variance  $\sigma^2$ is chosen such that 
the SNR $ = 10 \log_{10} (\bar s^2/ \sigma^2) = 15$dB, where 
$\bar s^2 = \frac 1 3 (|s_1|^2 +   |s_2|^2 +|s_3|^2)=0.4667$ denotes the average source power. The phase of each source $s_i \in \C$ is randomly generated 
from $Unif(0,2\pi)$ distribution. We consider  angular grid  $[\theta] =(-90, -85, \ldots, 80,85)$ with $5$ degree spacing. 
Thus the simulation set-up closely follows that of  \cite{gerstoft_etal:2015}. 
We compare results of regular Lasso ($=$ $M$-Lasso using LS-loss function) to the results of robust $M$-Lasso using Huber's loss function $\rho_{H,c}(\cdot)$ with threshold $c=
1.3774 $ corresponding to $q=0.85$.  To compare robustness of the methods, we  compute the estimates also for  corrupted data in which  magnitude of one measurement, $y_1$, is  scaled by a factor of $100$. 
To depict the $M$-Lasso solution  paths,  we compute the solution on a grid of 200  values in $(0,\lam_{\max})$ with equal spacings in the logarithmic scale, where $\lam_{max}$ denotes the smallest penalty value that shrinks all the coefficients of $M$-Lasso estimates to zero.

\begin{figure}[!t]
\centerline{ \includegraphics[width=0.25\textwidth]{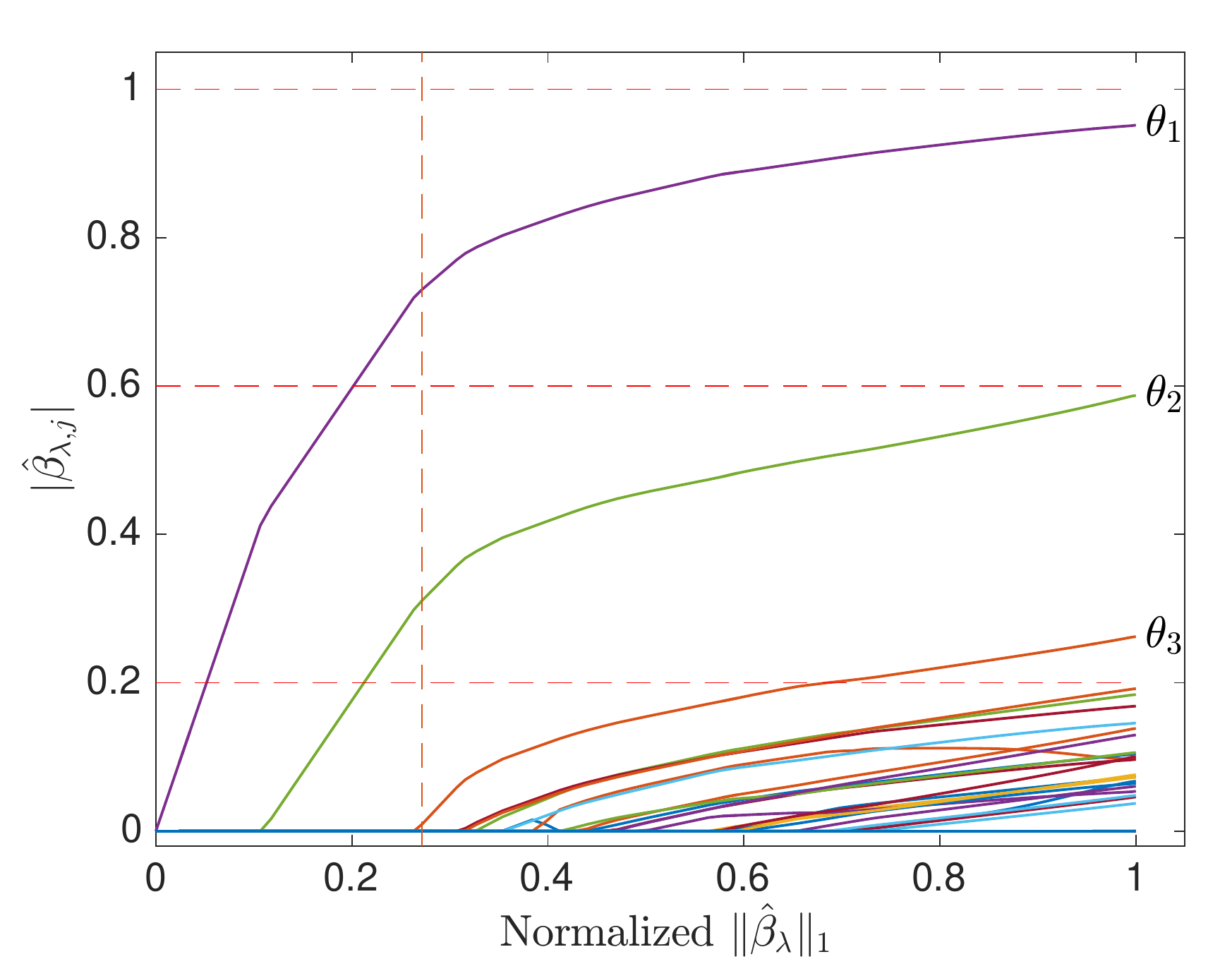}\includegraphics[width=0.25\textwidth]{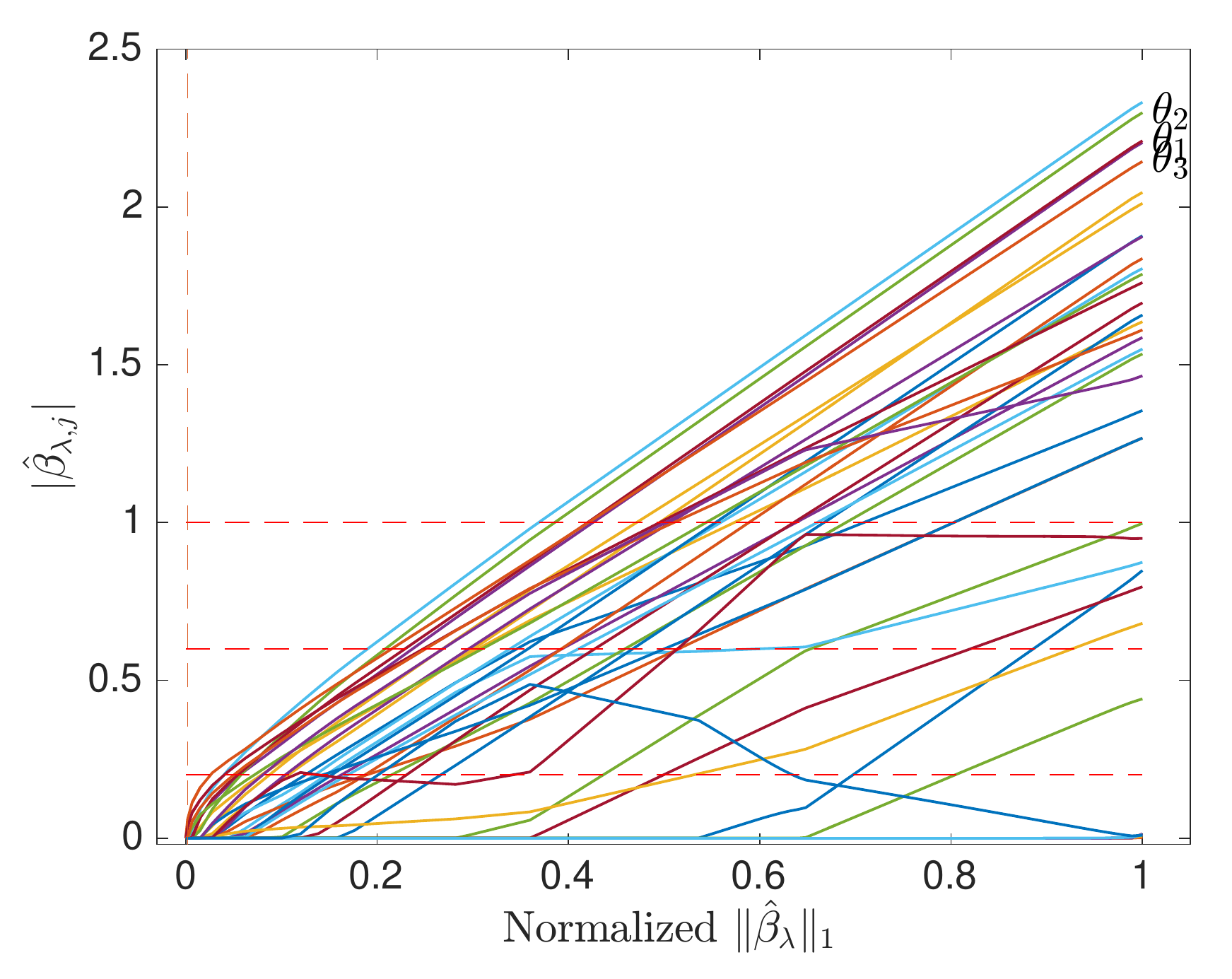}}
\vspace{0.1cm}
\centerline{ \includegraphics[width=0.25\textwidth]{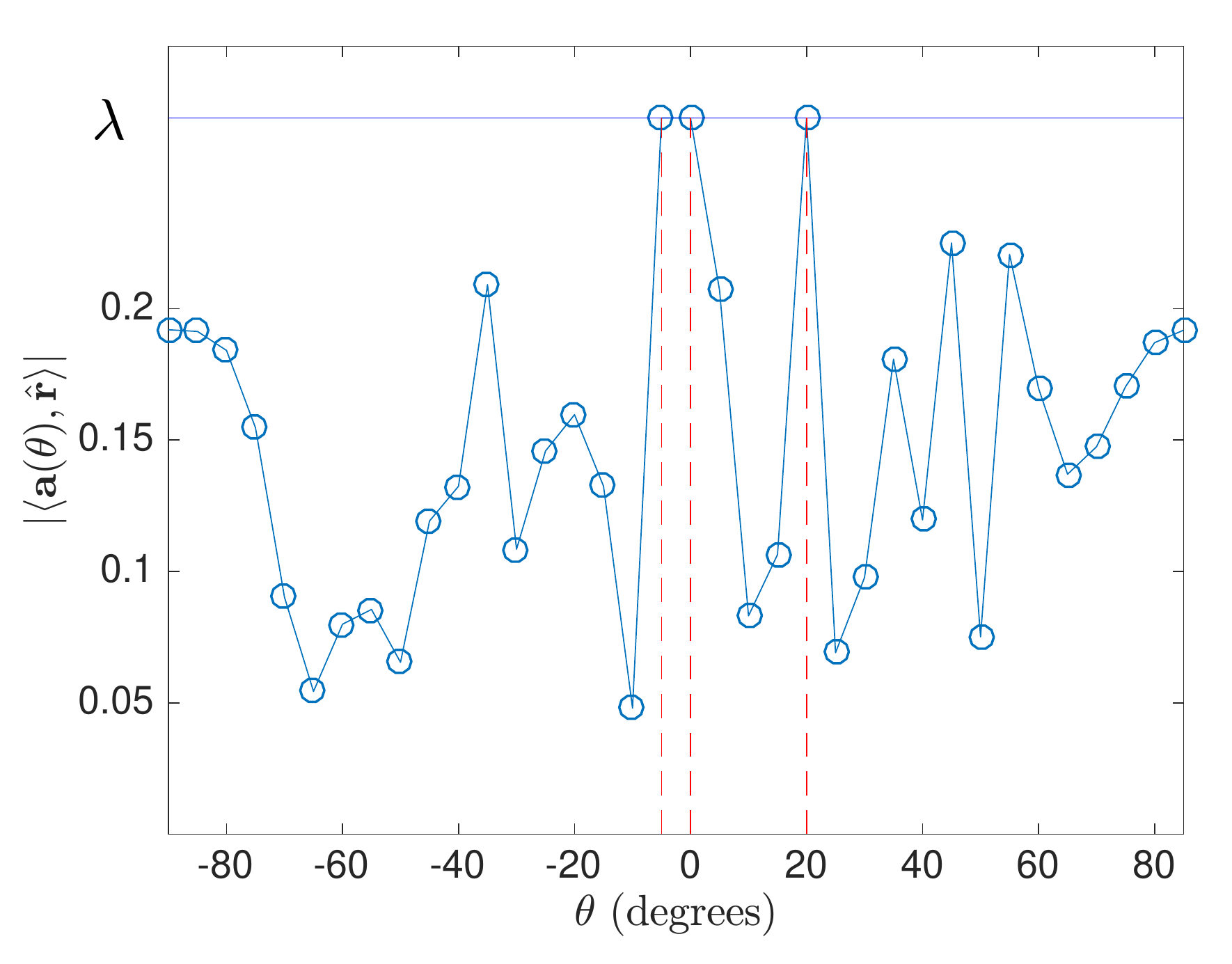}\includegraphics[width=0.246\textwidth]{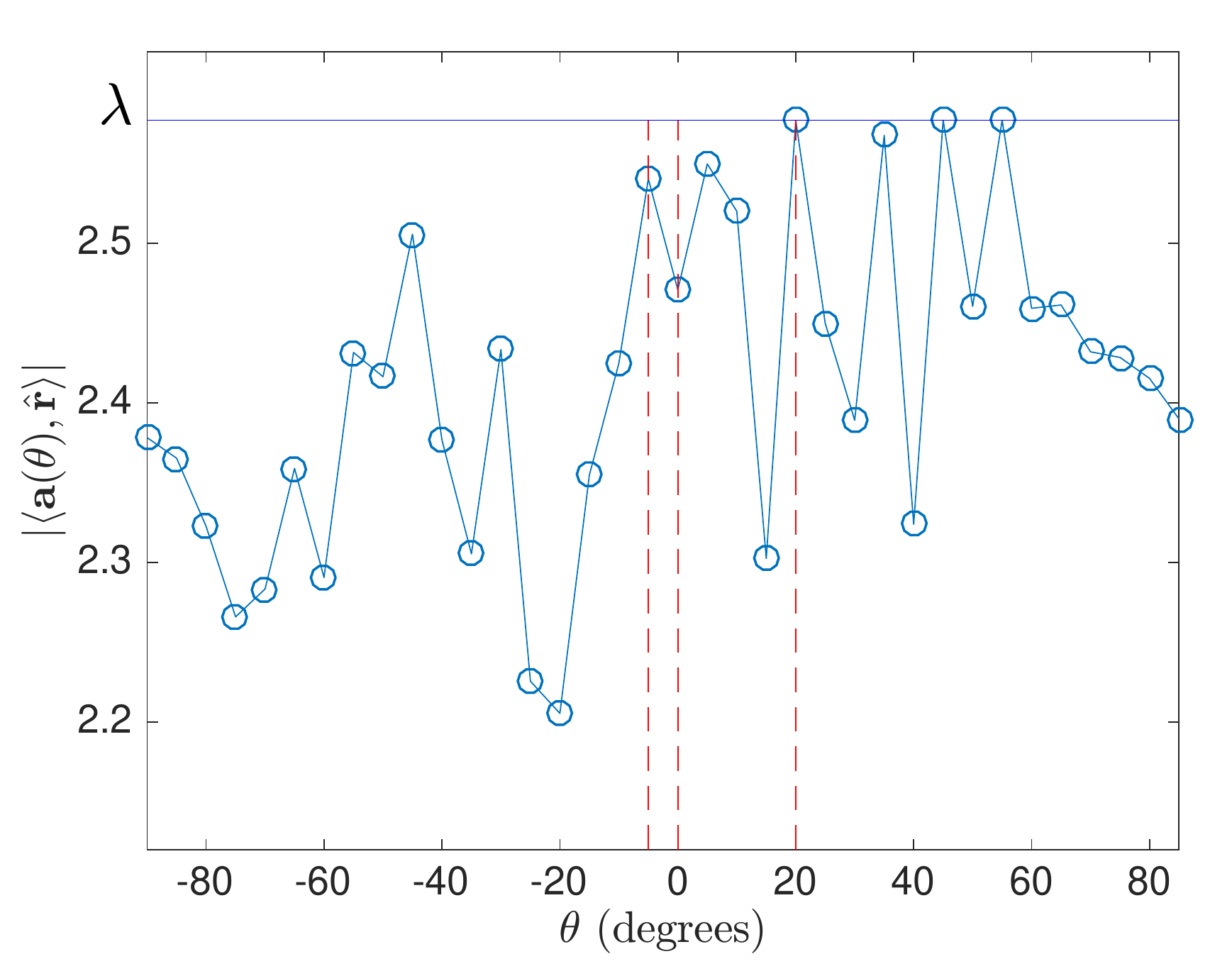}}
\centerline{ Lasso ($=$ $M$-Lasso using LS-loss function)} 
\vspace{0.1cm}
\centerline{ \includegraphics[width=0.25\textwidth]{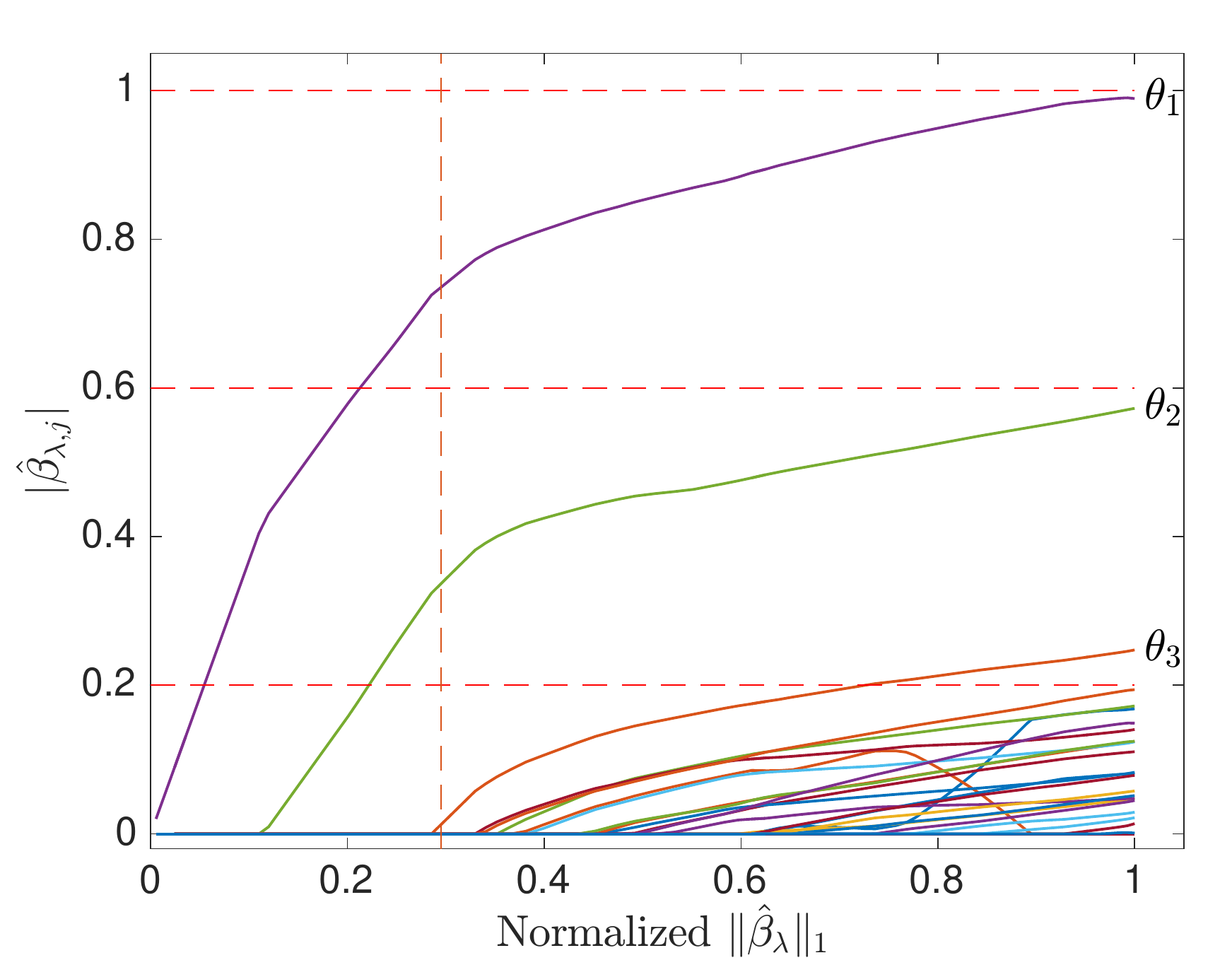}\includegraphics[width=0.242\textwidth]{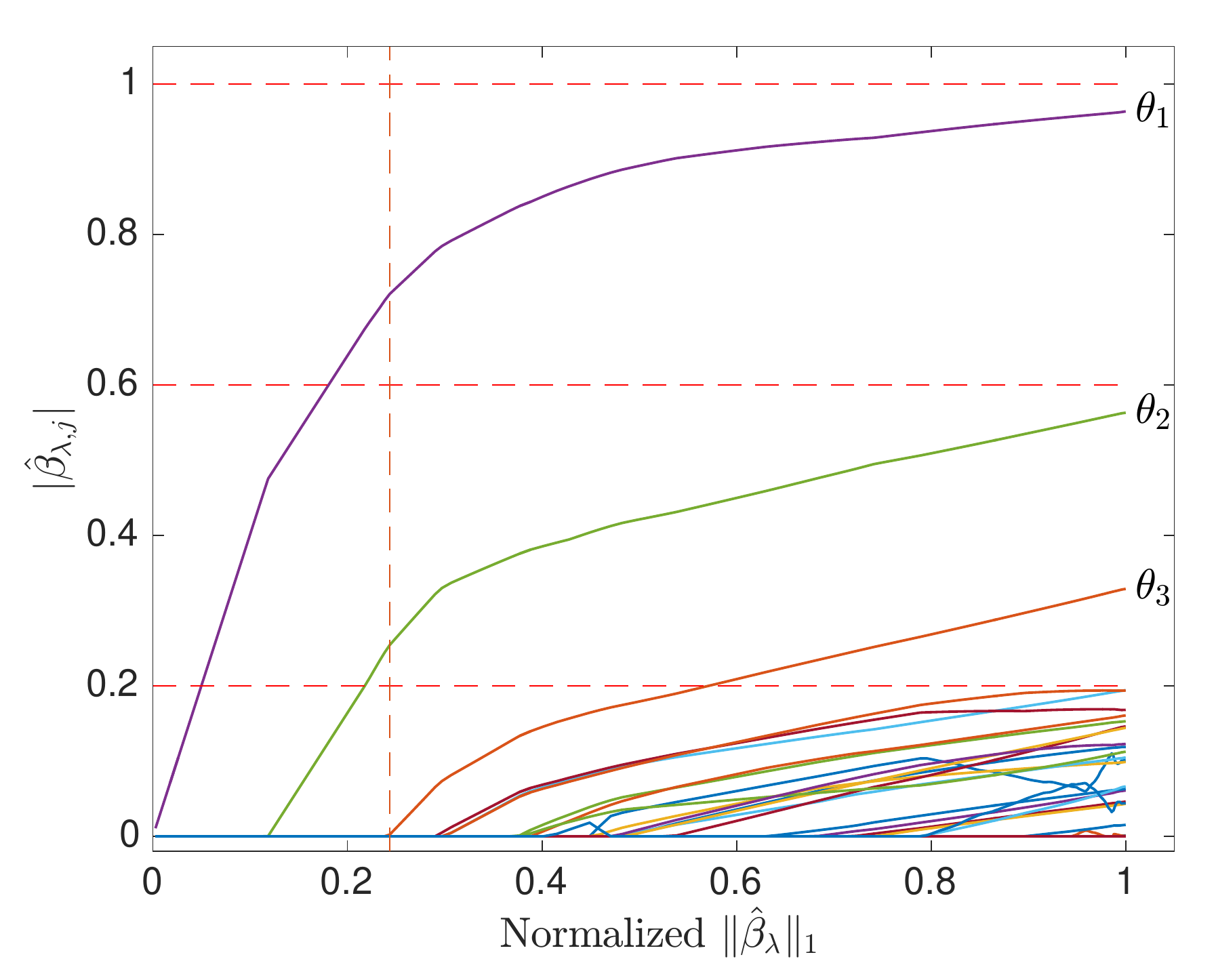}}
\vspace{0.1cm}
\centerline{ \includegraphics[width=0.25\textwidth]{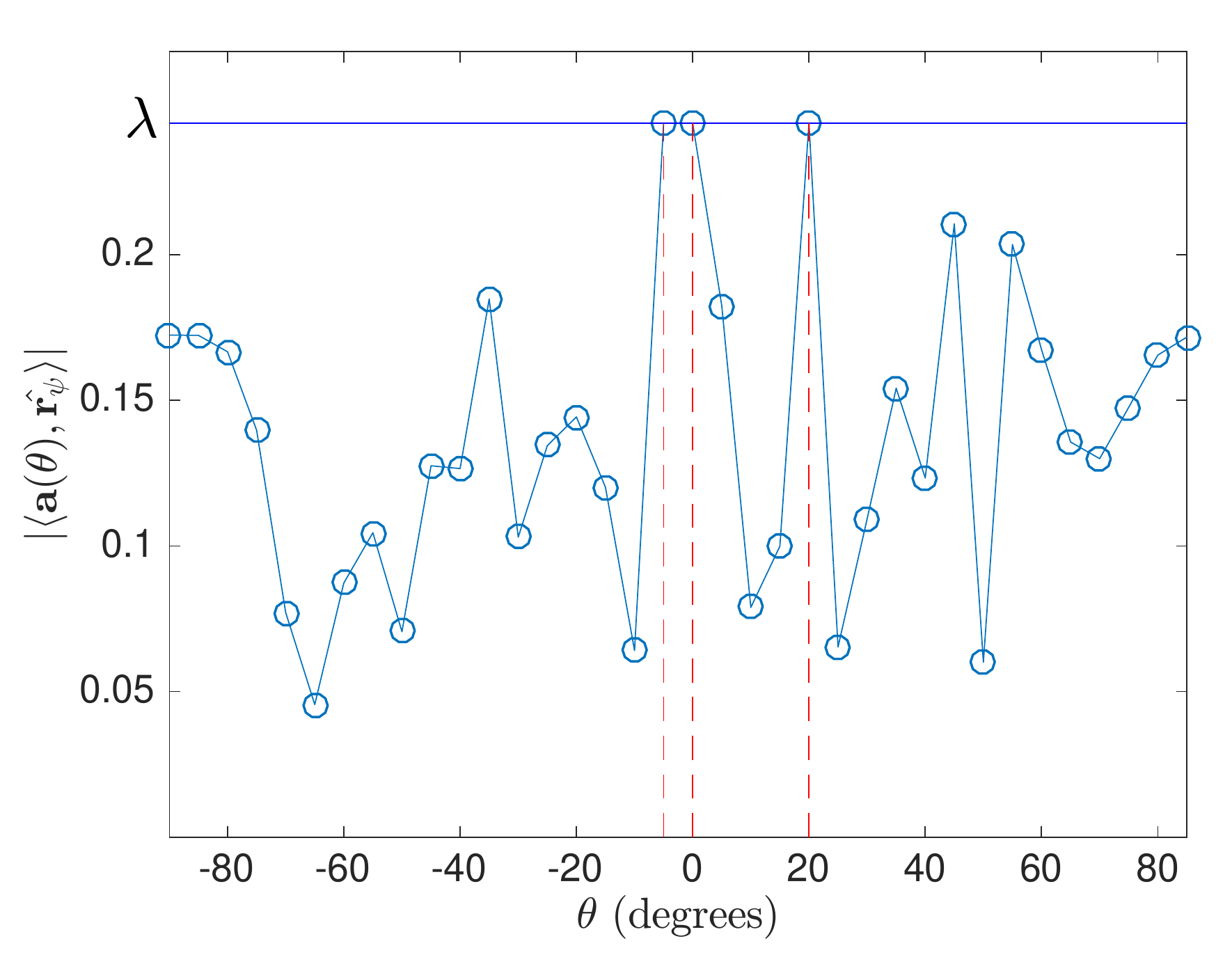}\includegraphics[width=0.25\textwidth]{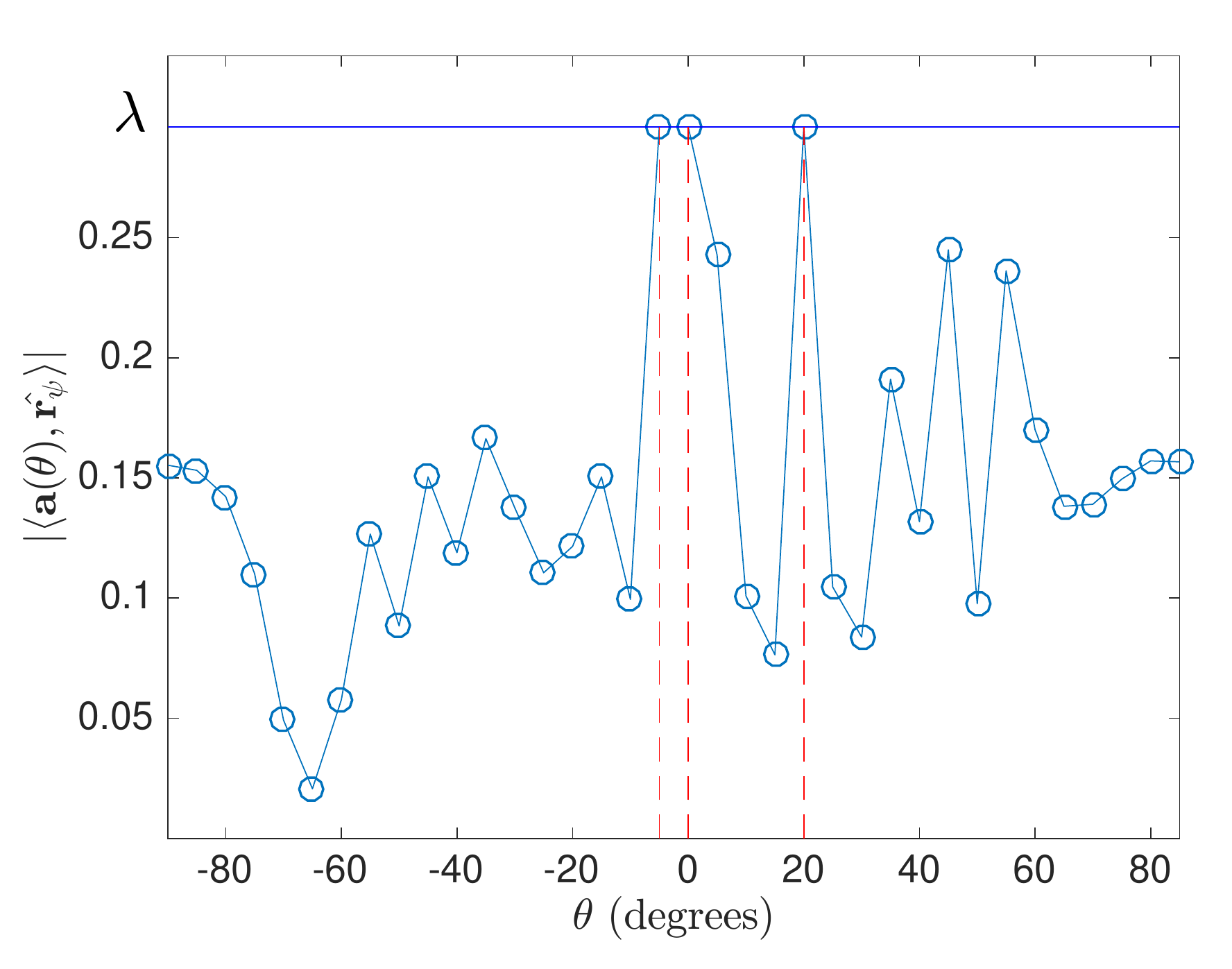}}
\centerline{ $M$-Lasso using Huber's loss function } 
\caption{Results for $M$-Lasso based on LS-loss and Huber's loss function. Left columns show results for the original data and the right column for 
the corrupted data. 
For both methods, first row shows the coefficient paths. 
The dotted vertical line 
identifies the solution $\hat \s_\lam$  which is used in the plots below and the dotted horizontal lines indicate the magnitudes $|\es_j|$ of the true sources. 
The second row depicts $| \langle \a(\theta_{(i)}) , \hat \r_\psi \rangle |$ on the angular grid  $[\theta]$. 
The horizontal line indicates the value of $\lam$ used and  the dotted vertical lines identify the true DoA's of the sources.}   \label{fig1}
\end{figure}

Left hand side column of Figure~\ref{fig1}  shows the results for the original data and the right  hand side column for the 
corrupted data. For each method, the upper row depicts 
the $M$-Lasso coefficient paths, i.e.,  the graphs of $|\hat \es_{\lambda,j}|$ for $j=1,\ldots,\pdim$ versus normalized $\| \hat \s_\lam \|$.  
The dotted vertical line identifies the solution $\hat \s_{\lam^*}$  which is  then used in the lower row plots. 
As can be seen the coefficient paths of Lasso and Huber's $M$-Lasso for original data are closely similar. For corrupted data, however, the Lasso coefficient paths completely change whereas the solution path for Huber's $M$-Lasso remains practically unaffected by the large outlier.  For original data both methods yield a solution $\hat \s_{\lam^*}$ 
that identify the true DoA's. However, for corrupted data, Lasso yields estimates $20^o$, $45^o$, and   $55^o$ degrees. Thus curiously, only the source $s_3$ (from DoA $\theta_3=20^o$) with lowest power (SNR) is correctly identified whereas the two higher power sources ($\theta_1=-5^o$ and $\theta_2=0^o$) are not. Huber's $M$-Lasso, however, correctly identifies the DoA's of the true sources as well as the order of the magnitudes.   For each method, the lower row  in Figure~\ref{fig1} plots $| \langle \a(\theta_{(i)}) , \hat \r_\psi \rangle |$  versus $\theta_{(i)}$ on the angular grid $[\theta]$. 
The horizontal line indicates the value $\lam^*$ (giving 3 nonzero coefficients) used and the dotted vertical lines identify the true DoA's of the sources.   These plots also illustrate that equations \eqref{eq:lam1}-\eqref{eq:lam2} hold, so the $M$-Lasso algorithm has correctly found the solutions to 
\eqref{eq:estim1} and \eqref{eq:estim2}.  To conclude, for original data, both methods produce similar plots and the same correct DoA estimates, but for corrupted data, 
only the robust Huber's $M$-Lasso provides reliable estimates. 

\vspace{7pt}
\section{Conclusions} \label{sec:concl} 

The robust $M$-Lasso estimates of regression and scale are  defined as solutions to generalized zero subgradient equations in the spirit of $M$-estimation. 
An explicit and efficient algorithm for computing the solution was proposed. 
The usefulness of  complex $M$-Lasso in DoA estimation of sources  with sensor arrays was illustrated 
using a simulated data set.   
Due to fast algorithm, we recommend using $M$-Lasso in practical big data applications due to its robustness properties. 

\end{document}